**Singularity-driven Second and Third Harmonic Generation in a ε-near-zero Nanolayer**

M. A. Vincenti[1], D. de Ceglia[1], A. Ciattoni[2], and M. Scalora[3]

[1]*AEgis Technologies Inc., 410 Jan Davis Dr., 35806 Huntsville AL –USA*

[2]*Consiglio Nazionale delle Ricerche, CNR-SPIN 67100, L'Aquila - Italy*

[3]*Charles M. Bowden Research Center, RDECOM, Redstone Arsenal, Alabama 35898-5000 - USA*

## Abstract

We show a new path to $\varepsilon$~0 materials without resorting to metal-based metamaterial composites. A medium that can be modeled using Lorentz oscillators usually displays $\varepsilon$=0 crossing points, e.g. $\varepsilon$=0 at $\lambda$~7$\mu$m and 21$\mu$m for $SiO_2$ and $CaF_2$, respectively. We show that a Lorentz medium yields a singularity-driven enhancement of the electric field followed by dramatic lowering of thresholds for a plethora of nonlinear optical phenomena. We illustrate the remarkable enhancement of second and third harmonic generation in a layer of $\varepsilon$~0 material 20nm thick, discuss the role of nonlinear surface sources in a realistic scenario where a 20nm thick $CaF_2$ nanolayer is excited at 21μm, and propose the realization of $\varepsilon$~0 materials in the visible range using dyes.

## Introduction

After the first demonstration of second harmonic generation (SHG) in 1961 [1], the enhancement of nonlinear processes has persisted as one of the main research activities in optics. A plethora of applications related to light generation at certain frequencies have been identified, and significant effort has been devoted to the study of harmonic generation from nanostructures having sizes that are not amenable to the use of phase-matching or quasi-phase-matching approaches. Several solutions have been proposed, all aiming at the enhancement of the electric field, that range from simple nanocavities [2, 3] to more complicated photonic crystals [4, 5]. The introduction of more sophisticated artificial structures and renewed interest in the excitation of surface waves [6] along the metal surface, and enhanced transmission [7], for example, have lead investigators to ask questions of a fundamental nature about the linear and nonlinear optical properties of metals and metal-based structures [8-11]. Surface plasmons are bounded waves at the surface between two media, and are usually accompanied by strong field enhancement in sub-wavelength regions [12, 13]. This observation has led to the exploration of nonlinear optical processes in structures that operate in the enhanced transmission regime: for example, studies of hole- and slit-arrays filled with a nonlinear medium have demonstrated that a boost in the linear response coincides with the enhancement of SHG and/or third harmonic generation (THG) [14, 15].

Centrosymmetric materials such as metals do not possess intrinsic, dipolar, quadratic nonlinear terms. Nevertheless, metals display an effective second order nonlinearity that arises from a combination of symmetry breaking at the surface, magnetic dipoles (Lorentz force), inner core electrons, convective nonlinear sources and electron gas pressure [16]. They also possess a relatively large, third order nonlinearity that together with effective second order nonlinear sources introduce a non negligible contribution in the generated signals [8, 15, 17]. The detailed study and evaluation of these individual contributions has been made possible by the recent development of a dynamical model [16] for a nearly free electron gas where bound charges (or inner core electrons) also play a role in the determination of linear and nonlinear optical properties of metals [8, 17]. The model does not make any a priori assumptions about the relative weights of surface and volume sources, and has been shown to adequately predict SH and TH conversion efficiencies in centrosymmetric [16] and semiconductor materials [18], where only bound electrons are present.

Quite recently epsilon-near-zero ($\varepsilon$~0) materials have been investigated for their peculiar linear [19, 20] and nonlinear optical properties [21-24]. In particular, studies in metal-dielectric composites having effective $\varepsilon$~0 have demonstrated that the extreme environment leads to significantly enhanced SH conversion efficiencies [22] and to peculiar memory and bistability features [23]. In fact, the longitudinal component of the (TM-polarized) electric field becomes singular anytime a material exhibits permittivity values close to zero due to the requirement that the longitudinal component of the displacement field be continuous [19-24].

In this manuscript we present the results of a theoretical investigation of SHG and THG from a 20nm-thick layer of an ideal uniform material having $\varepsilon$~0 in the visible range. Harmonic generation is examined also by quantifying and comparing the role that surface and volume terms play with respect to intrinsic second and third order nonlinear susceptibilities, i.e. $\chi^{(2)}$ and $\chi^{(3)}$. This evaluation becomes necessary because near the $\varepsilon$=0 condition the longitudinal electric field may be enhanced from several hundreds to many thousands of times, resulting in strong nonlinear surface (quadrupole-like) and volume (magnetic dipole, electric quadrupole) sources beyond the contributions of the intrinsic nonlinearities. The nonlinear dynamics of bound electrons of ε~0 materials is thus modeled by explicitly including electric and magnetic forces as outlined in references [15, 16, 18]. Normalized conversion efficiencies of order $10^{-5}$ and $10^{-7}$ are predicted for SHG and THG, respectively, when a uniform layer only 20nm thick having $\chi^{(2)}$=20pm/V and $\chi^{(3)}=10^{-20}(m/V)^2$ is illuminated with TM-polarized light and peak intensity of 40MW/cm$^2$. A more realistic scenario is also examined, where a 20nm thick CaF2 ($\chi^{(2)}$=0 and $\chi^{(3)}\sim10^{-21}(m/V)^2$) layer is illuminated at 21μm revealing how the generated harmonics resulting from symmetry breaking at the surface, magnetic dipole, and electric quadrupole contributions are significantly higher when compared with metals.

**ε~0 condition and electric field enhancement**

As may easily be ascertained, any Lorentz-type material has readily accessible ε~0 regions [25], while most transparent materials materials display far-IR and deep UV Lorentz-type resonances. On the other hand, common semiconductors like GaAs, GaP, and Si display absorption resonances in the visible and UV ranges, have ε=0 crossing points near 250nm and 100nm, respectively [25]. Although this spectral region is certainly of interest, semiconductors usually present special challenges, and so for illustration purposes we explore alternative materials. For instance, fluorides (LiF, $CaF_2$ or $MgF_2$) and oxides like $SiO_2$ have ε=0 crossing points in the 7-40μm range [25], where they display metallic behavior. Another peculiarity of typical Lorentz systems is that absorption tends to be somewhat abated at the short-wavelength crossing point, where for example $CaF_2$ and $SiO_2$ exhibit comparatively smaller absorption than semiconductors at their short-wavelength ε=0 crossing points, resulting in more favorable field enhancement conditions compared to semiconductors. For simplicity material X is modeled using a single species of resonant Lorentz oscillators that yields the following complex dielectric function:

$$\varepsilon_x(\omega) = 1 - \frac{\omega_p^2}{\omega^2 - \omega_0^2 + i\gamma\omega} \quad (1)$$

where the plasma frequency $\omega_p=2\omega_r$, damping $\gamma=10^{-2}\omega_r$, the resonance frequency $\omega_0=0.5\omega_r$, and the reference frequency $\omega_r=2\pi c/1\mu m$. These parameters produce a strong resonance at 2μm (Fig.1 (b)) and a ε=0 crossing point that naturally displays limited absorption at λ~485nm. That is to say, absorption is very much a part of the system that we study, but tune the field off-resonance, in a spectral region where absorption tends to subside. In order to give our results a more realistic slant near the crossing point we have chosen γ so that the resulting peak dielectric and index values are similar to those displayed by $CaF_2$, with the exception that they are shifted toward the visible range. The main effect of choosing larger (smaller) γ is to reduce (increase) the maximum field amplitude in the material and, consequently, the strength of the nonlinear optical interaction.

The simple geometry of the structure under investigation is depicted in Fig.1 (a): a d=20nm thick layer of material X is illuminated with TM-polarized light. Incident wavelength and angle are varied to explore the linear properties of the slab and to evaluate the electric field enhancement inside the medium. As detailed elsewhere [19, 21], at the zero-crossing point linear transmission at normal incidence is nearly 100% (if absorption is neglected) and 0% for all other incident angles. If a material with non-zero damping is considered the transition between 100% and 0% transmission is more gradual: steepness of the transition thus also depends on material absorption.

We now focus on the electric field enhancement of this structure in the vicinity of the zero-crossing point for different incident angles. Choosing a 20nm layer has the advantage of removing cavity resonances that can interfere with the process and complicate the analysis [21-24]. The linear optical properties of the structure were calculated using a standard transfer matrix method (TMM) [26]. As Fig.2 (a) shows, the longitudinal electric field intensity is amplified approximately 400 times relative to the incident field.

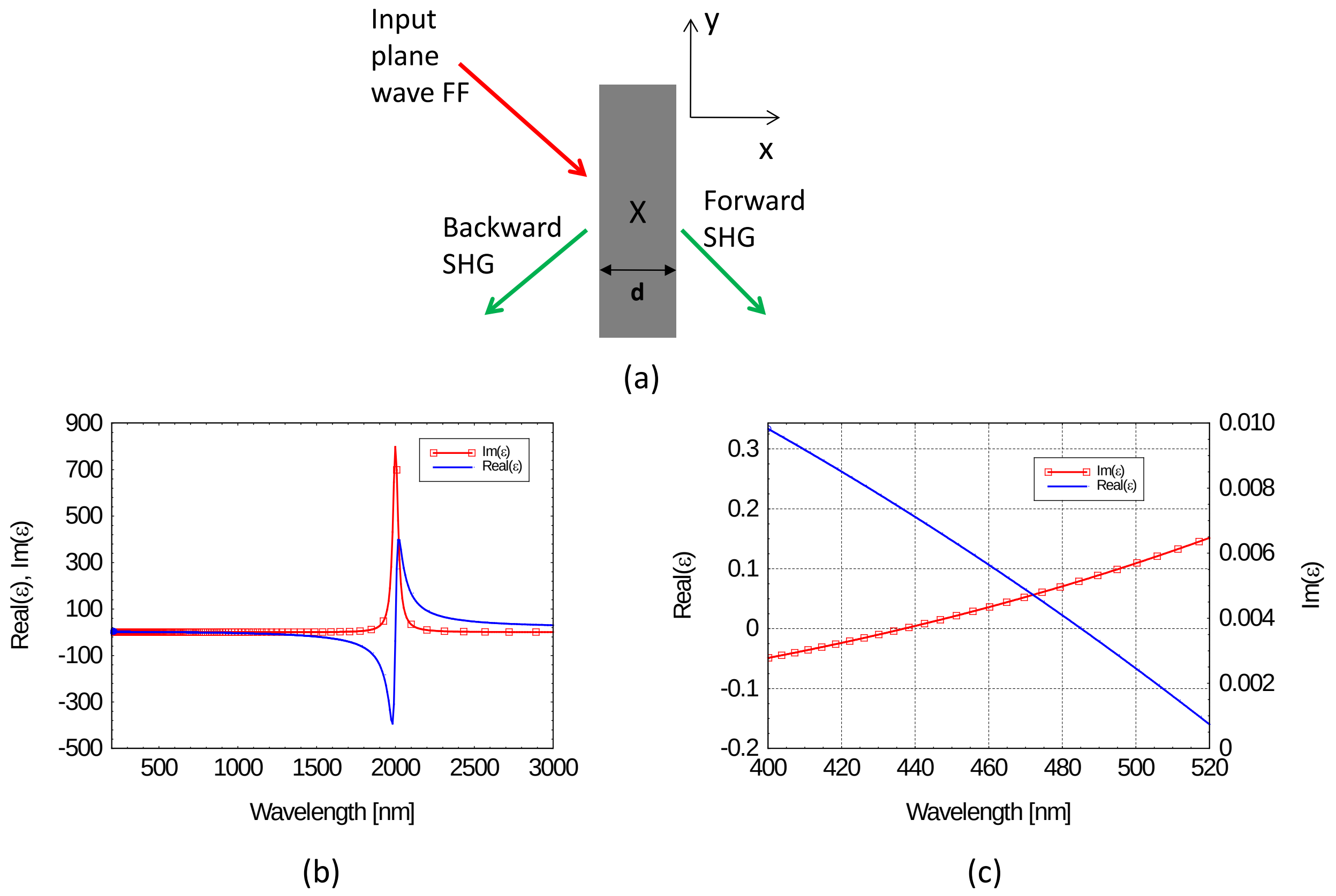

**Fig.1:** (a) A pump (FF) is incident on a 20nm-thick layer of uniform material composed of Lorentz oscillators resonant at 2μm. (b) Material dispersion and (c) detail of the ε=0 crossing region. As shown in (b) absorption is present in the system but the ε~0 condition occurs far from resonance.

As an example, we note that for $\gamma=10^{-4}$ the field intensity is amplified approximately by a factor of 35000 compared to the incident field intensity, confirming the singularity-driven nature of the field near the crossing point. In the absence of other mechanisms (the thin layer is not resonant) the amplification of the field comes solely as a result of the fact that the longitudinal component of the displacement field must be continuous.

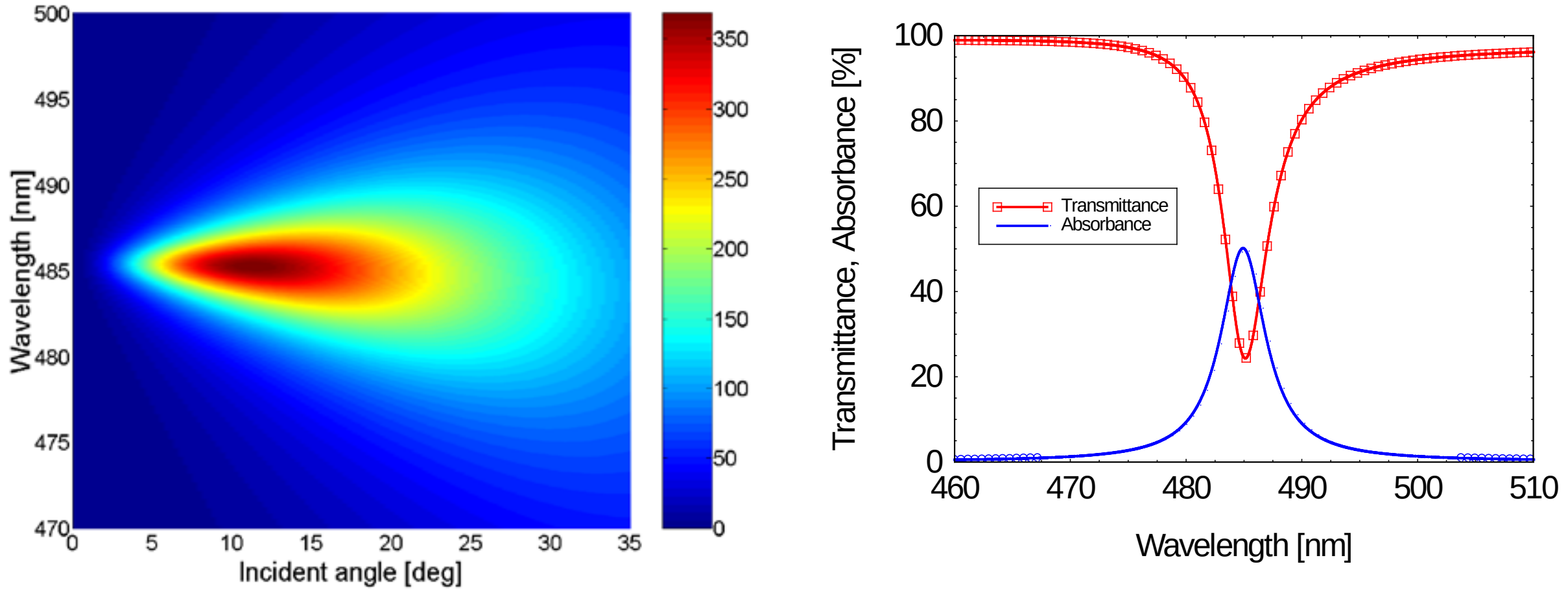

(a) (b)

**Fig.2:** (a) Maximum intensity recorded inside the 20nm layer as a function of wavelength and incident angle, normalized with respect to the incident intensity. The intensity is distributed uniformly inside the layer with an amplification factor of ~400. (b) Transmission and absorption vs. wavelength when the field is incident at 12º. The peculiarity here is that maximum amplification occurs near the bottom of the transmission curve, (i.e. a rudimentary gap that forms when $\varepsilon$ and $\mu$ have opposite signs) where absorption is also a maximum: nearly 50% of the incident light is absorbed.

The excitation of a large longitudinal electric field, highlighted in reference [22] in the context of SHG from metal-dielectric composites, is associated with a transversely moving wave, and may be interpreted simply as a guided mode resonance inside the slab, i.e. a polariton in a more common vernacular. The formation of this mode is apparent from the analysis of transmission and absorption properties of the structure, where nanometer-size resonant features appear at relatively small angles. In Fig.2 (b) we also show transmission and absorption for a wave incident at 12º. It is evident that in this system overall absorption is not at all negligible, in spite of the fact that Im($\varepsilon$) is relatively small, because absorption is proportional to the product Im($\varepsilon$)$|\mathbf{E}|^2$. Therefore, increasing the imaginary part of the dielectric function must be paired with the localization properties of the field to properly assess absorption. We reiterate that no cavity resonances are excited, and that the enhancement of the field shown in Fig.2 (a) and (b) comes solely as a result of approaching the $\varepsilon$=0 conditions.

**Nonlinear Results for a representative ε~0 medium having bulk nonlinearities**

The nonlinear calculations were performed using three different methods: (i) finite difference time domain and (ii) fast Fourier transform spectral techniques, which were both used to also study the effect of surface, magnetic and quadrupolar sources, and (iii) Comsol Multiphysics [27] for the CW regime. Methods (i) and (ii) consist of a classical oscillator model under the action of internal forces (damping, linear and nonlinear restoring forces) and external forces due to the applied fields. The basic equations of motion are derived in details in reference [16] to describe second and third harmonic generation from the surface and volume of metal-based nanostructures and nanocavities. In references [15] and [18] the technique was extended to include the effects of second and third order nonlinearities of materials like GaP and GaAs, which have anisotropic $\chi^{(2)}$ and $\chi^{(3)}$ tensors. So the methods are quite general, and can handle vector field propagation in two dimensions and time.

In Figs.3 (a) and (b) we report conversion efficiencies for SHG and THG vs. incident angle and wavelength for the idealized case. We remark that harmonic generation follows the intensity enhancement profile of Fig. 2 (a), reaching efficiencies of order $10^{-5}$ for SHG and $10^{-7}$ for THG when $\gamma=10^{-2}$, which increase to $10^{-3}$ and $10^{-5}$ respectively, when $\gamma=10^{-4}$. In order to provide context for these conversion efficiencies we compared these results with the predictions of SHG and THG made in reference [15] for a resonant metal grating 100nm thick, filled with GaAs. In that case, using intensities of order 2GW/cm$^2$ (fifty times larger than presently), $\chi^{(2)}$=100pm/V (five times larger) for GaAs, and $\chi^{(3)}$=$10^{-15}$(m/V)$^2$ (five orders of magnitudes larger!) in the bulk metal [17], the best conversion efficiencies attainable were $10^{-6}$ and $10^{-4}$ for SHG and THG, respectively, which are arguably comparable to the conversion efficiencies that we report for a far simpler structure consisting of a uniform nanolayer of ε~0 materials. One could then also reasonably argue that there may be significant advantages and new opportunities in exploiting the nonlinear optical properties of ε~0 materials.

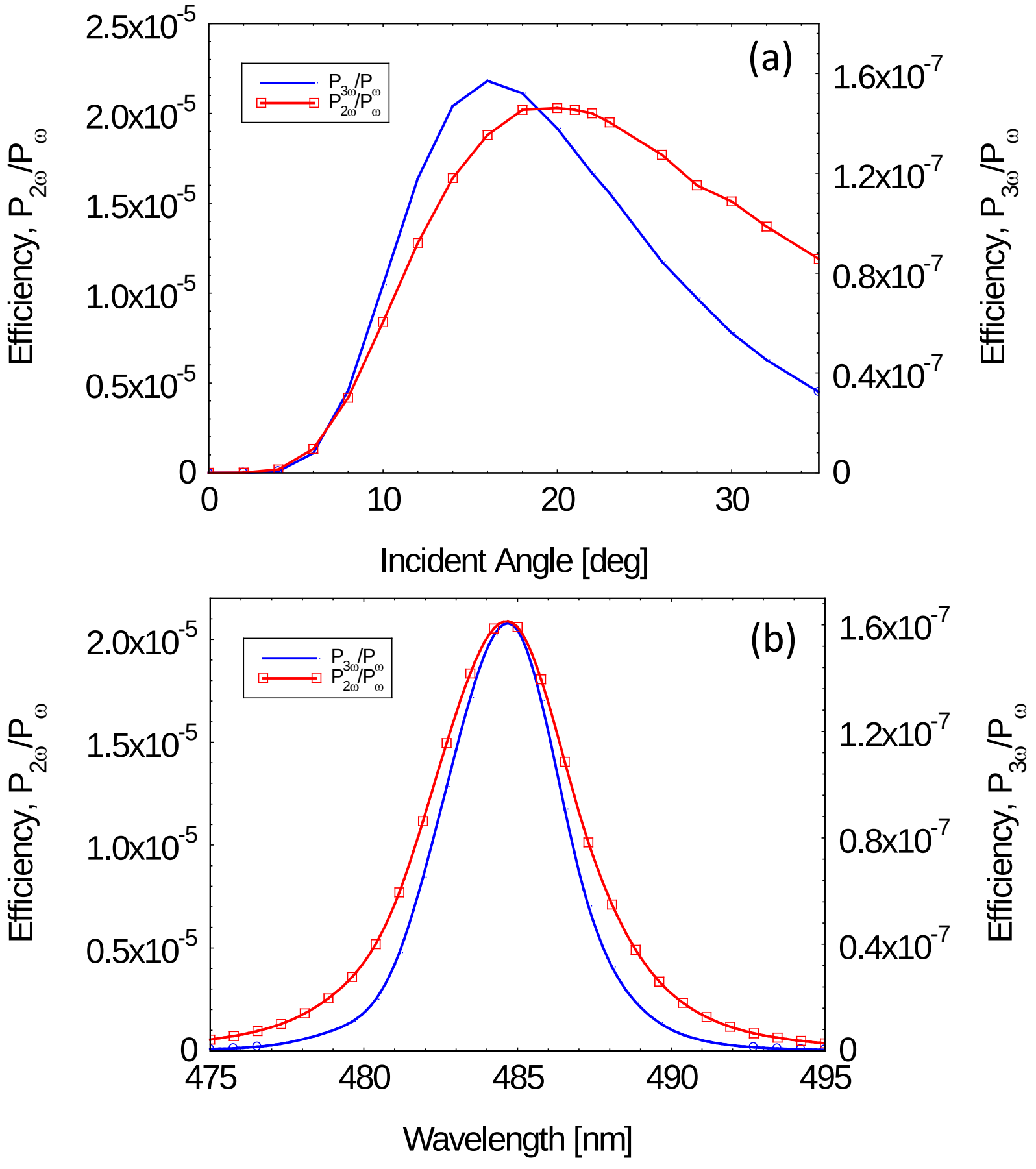

**Fig.3:** (a) SH and TH conversion efficiencies vs. incident angle for the 20nm layer of Fig.1 (a). (b) Emitted spectra for SH and TH fields centered at 242nm and 161nm at 18°. $\chi^{(2)}$=20pm/V and $\chi^{(3)}$=$10^{-20}$(m/V)$^2$.

## Nonlinear Results for $CaF_2$ nanolayer

The kinds of field enhancement and jumps that we are discussing occur over very short distances and naturally raise the possibility that harmonic generation from surface, magnetic dipole and quadrupolar sources may come into play in centrosymmetric materials, or media with small $\chi^{(2)}$ and/or $\chi^{(3)}$. In modeling the medium care must thus be taken to include and assess these effects. For this reason we use the model briefly mentioned above outlined in references [15, 16, 18].

The question that one now may ask is how to best achieve conditions for experimental verification of enhanced harmonic generation near the $\varepsilon=0$ points. Although semiconductors like Si and GaP display relatively low absorption at $\lambda\sim100$nm, fluorides and glasses may be directly addressed using appropriate IR sources. As a practical example we calculated the maximum intensity for the longitudinal electric field inside a 20nm thick layer of $CaF_2$, and in Fig.4 we report the results using data found in reference [25]. The electric field intensity is amplified approximately 32 times relative to the incident value. Although it is smaller compared to the hypothetical material cited in Fig.2, one can easily see that the main features are nearly identical. Clearly absorption plays an important role in that it shifts the optimal incident angle to higher values, rescales maximum field enhancement, and at the same time broadens the useful bandwidth. Similarly to LiF and Si, $CaF_2$ has a cubic crystal structure, with $\chi^{(2)}\sim0$ and $\chi^{(3)}\sim10^{-21}$ $m^2/V^2$ [28, 29]. A zero $\chi^{(2)}$ and a reduction of local field intensity rescales SH and TH conversion efficiencies down to approximately $10^{-9}$. This reduction notwithstanding, the predicted SH conversion efficiency is still some two to three orders of magnitude larger than efficiencies recorded when metal surfaces are illuminated with similar intensities [30], thus making ε~0 materials clearly competitive with nonlinear plasmonic nanostructures.

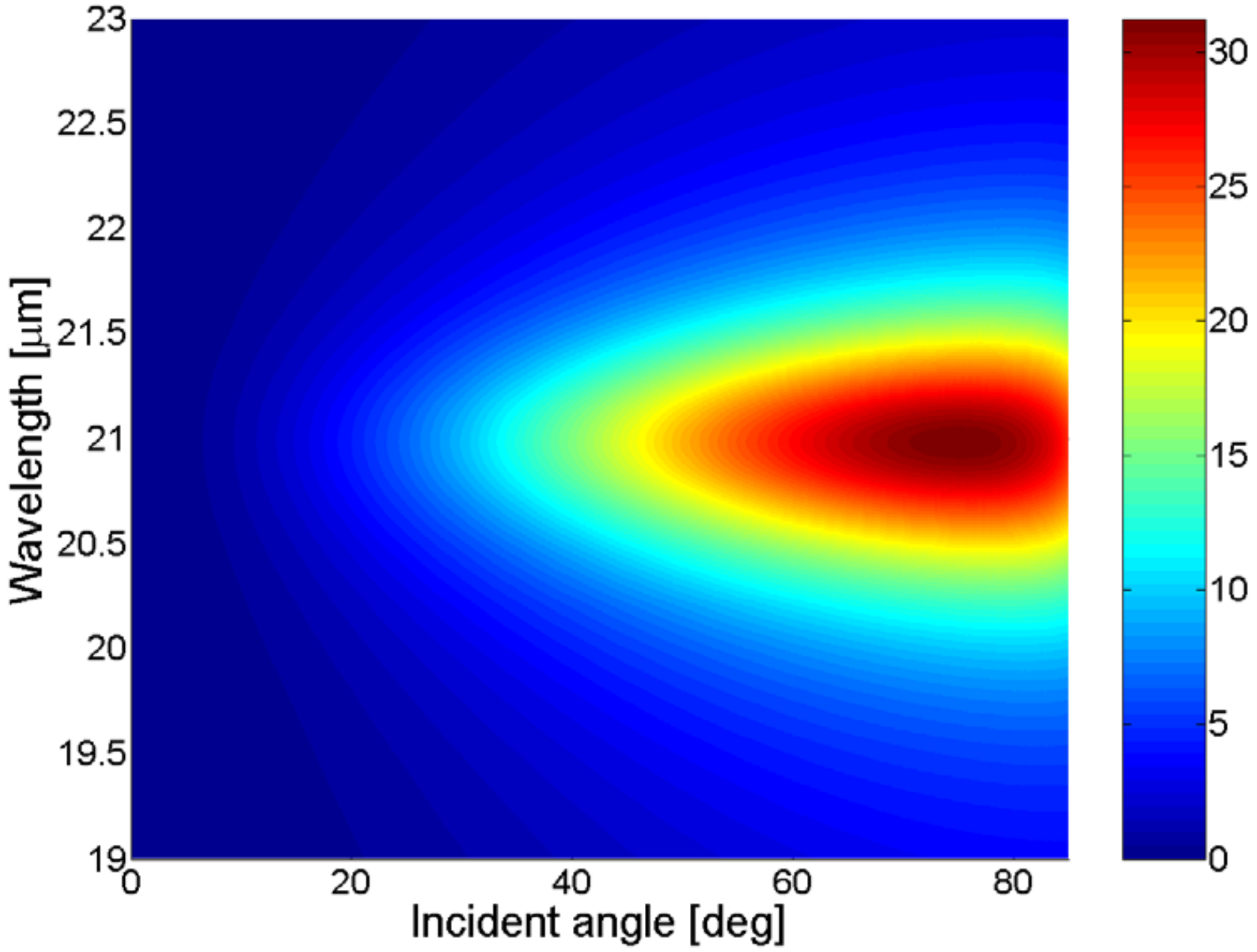

**Fig.4:** Maximum field intensity calculated inside a 20nm $CaF_2$ layer as a function of wavelength and incident angle, normalized with respect to the incident field intensity. The intensity is distributed uniformly inside the layer with an amplification factor of ~32. We used actual data for $CaF_2$ found in reference [25].

$CaF_2$ may be pumped directly at a wavelength of approximately 21μm with a tunable laser source similar to that described in reference [31]. At the same time, some engineering may be required to improve field enhancement and as an alternative it may be possible to approach the problem in the visible or IR ranges using glasses or liquids doped with dyes [32], such as Coumarin C500, Rhodamine 6G, whose optical loss compensation effects have been recently demonstrated [33, 34]. Typical visible and near-IR dyes that are commercially available have near-Lorentzian absorption profiles and may be properly engineered to reduce dramatically optical losses in the region where the real part of the effective permittivity crosses the zero [35, 36].

## Conclusions

In summary we have presented a new type of $\varepsilon$~0 material whose topology does not require a metal-dielectric metamaterial infrastructure. Using this simple approach we are able to extract a unique type of phenomenology that lowers the threshold for a wealth of nonlinear optical phenomena, from harmonic generation to optical bistability and switching, and from quantum optical interaction like stimulated Raman scattering to soliton formation and high field ionization, to name a few. We have investigated harmonic generation in media with relatively small $\chi^{(2)}$ and $\chi^{(3)}$ (20pm/V and $10^{-20}m^2/V$, respectively) and in centrosymmetric $CaF_2$, and predicted much improved SHG and THG conversion efficiencies compared with plasmonic nanostructures with only a few $MW/cm^2$ of input power in the presence of realistic absorption. Strong electric field enhancement inside the nanolayer is due to the continuity of the longitudinal component of the displacement field which in $\varepsilon$~0 materials causes the relative component of the electric field to approach singular behavior. Our calculations also show that symmetry breaking and nonlinear volume sources arising from magnetic dipoles and electric quadrupoles have to be included to model centrosymmetric materials like $CaF_2$. Finally, it is clear that while some molecular and material engineering may be required for experimental verification, it should also be apparent that an equally clear and relatively straightforward path now exists to the development of exotic and extreme nonlinear optical phenomena in the $KW/cm^2$ range.